\documentstyle[12pt,epsf]{article}

\newcommand{\wt}{\widetilde}
\newcommand{\vu}{\vec u}
\newcommand{\tu}{\widetilde u}
\newcommand{\tv}{\widetilde v}
\newcommand{\MM}{{\cal M}}

\newcommand{\BB}{{\cal B}}

\newcommand{\be}{\begin{equation}}
\newcommand{\ee}{\end{equation}}
\newcommand{\ben}{\begin{eqnarray}\displaystyle}
\newcommand{\een}{\end{eqnarray}}
\newcommand{\refb}[1]{(\ref{#1})}

\begin{document}

\begin{titlepage}

\title{Orientifold Limit of F-Theory Vacua\footnote{Talk given at the
Trieste conference on Duality Symmetries and at Strings 97}}

\author{Ashoke Sen\footnote{On leave of absence from Tata Institute of
Fundamental Research, Homi Bhabha Road, Bombay 400005, India}
\footnote{E-mail: sen@mri.ernet.in, sen@theory.tifr.res.in}\\ ~ \\
Mehta Research Institute of Mathematics \\
and Mathematical Physics \\
Chhatnag Road, Jhusi, Allahabad 221506, INDIA}

\maketitle

\begin{abstract}
We show how F-theory on a Calabi-Yau $(n+1)$-fold, in appropriate limit,
can be identified as an orientifold of type IIB string theory
compactified on a Calabi-Yau $n$-fold.
\end{abstract}

\vfill

\noindent\vbox{\hbox{hep-th/9709159}\hbox{MRI/PHY/P970924}}

\end{titlepage}

Orientifolds and F-theory are two apparently different ways of 
compactifying type IIB string theory.
In this talk we shall explore the relationship between these two
different classes of type IIB compactification. In particular, we
shall show how $F$-theory on a Calabi-Yau $(n+1)$-fold, in
appropriate limit, reduces to an orientifold of type IIB string
theory compactified on a Calabi-Yau $n$-fold. This talk will be
based mainly on ref.\cite{FORI} and also partially on
refs.\cite{GIMON2,SETHI}. All other relevant references can
be found in these papers.

We begin with
some facts about type IIB string theory.
Massless bosonic fields in type IIB theory come from two sectors.
The Neveu-Schwarz $-$ Neveu-Schwarz (NS) sector contributes the
graviton
$G_{\mu\nu}$, the anti-symmetric tensor field $B_{\mu\nu}$, and
the dilaton $\Phi$. The Ramond-Ramond (RR) sector contributes
another scalar $a$, sometimes called the axion, another
anti-symmetric tensor field $B'_{\mu\nu}$, and a rank four
anti-symmetric tensor field $D_{\mu\nu\rho\sigma}$ with self-dual
field strength.
We define
\be \label{e1}
\lambda \equiv a + i e^{-\Phi}\, .
\ee
This theory has two
perturbatively realised $Z_2$ symmetries. The first one $-$ denoted
by $(-1)^{F_L}$ where $F_L$ is the contribution to the space-time
fermion number from the left moving sector of the 
world sheet $-$ changes sign of $a$, $B'_{\mu\nu}$,
$D_{\mu\nu\rho\sigma}$, leaving the other massless bosonic fields
unchanged. The second one $-$ the world-sheet parity
transformation $\Omega$ $-$  changes the sign of
$B_{\mu\nu}$, $a$ and $D_{\mu\nu\rho\sigma}$.
Besides these two symmetries which are valid order by order in
perturbation theory, this theory also has a
conjectured non-perturbative symmetry\cite{HT}, 
known as S-duality, under
which
\be \label{e2}
\lambda \to {p\lambda + q\over r\lambda + s}, \qquad
\pmatrix{B_{\mu\nu}\cr B'_{\mu\nu}}\to \pmatrix{p & q \cr r & s}
\pmatrix{B_{\mu\nu}\cr B'_{\mu\nu}}\, .
\ee
Here $p,q,r,s$ are integers satisfying $ps-qr=1$.
We shall denote by $S$ and $T$ the following specific
SL(2,Z) transformations:
\be \label{e3}
S = \pmatrix{0 & 1\cr -1 & 0}\, , \qquad T=\pmatrix{ 1 & 1\cr 0 & 1}\, .
\ee
Studying the action of the various transformations
on the massless fields, we can identify the
discrete symmetry transformation $(-1)^{F_L}\cdot\Omega$ with the
SL(2,Z) transformation:
\be \label{e4}
\pmatrix{-1 & 0\cr 0 & -1}\, .
\ee

Orientifolds are orbifolds of (compactified) type IIB theory,
where the orbifolding group involves the world-sheet parity
transformation $\Omega$. In this talk we shall focus our
attention on
a class of orientifolds of type IIB on Calabi-Yau manifolds
defined as follows. Let 
$\MM_n$ be  a Calabi-Yau $n$-fold ($n$ complex dimensional
manifold), and
$\sigma$ be a $Z_2$ symmetry of $\MM_n$ such that
\begin{itemize}
\item it reverses the sign of the holomorphic $n$-form on
$\MM_n$ and, 
\item its fixed point sets form a submanifold of complex codimension
1.
\end{itemize}
We now
consider the orientifold:
\be \label{e5}
IIB \quad \hbox{on} \quad \MM_n\times
R^{9-2n,1}/(-1)^{F_L}\cdot\Omega\cdot\sigma\, ,
\ee
where $R^{9-2n,1}$ denotes the $(10-2n)$ dimensional Minkowski
space. The transformation $(-1)^{F_L}\cdot\Omega\cdot\sigma$
preserves half of the space-time supersymmetry. The fixed point set
on $\MM_n\times R^{9-2n,1}$ under $\sigma$ is of real codimension two in
the (9+1) dimensional space-time, and are known as 
orientifold seven planes. This is known to carry
$-4$ units of RR charge. Thus 
the SL(2,Z) monodromy along a closed contour around this
orientifold plane is $-T^{-4}$. Here
$-1$ is the  effect of the monodromy $(-1)^{F_L}\cdot \Omega$,
and $T^{-4}$ is the effect of RR charge carried by the orientifold
plane.
Since there is no non-compact direction transverse to the
orientifold plane, 
this monodromy must be cancelled by placing Dirichlet 7-branes
(D7-branes)
along appropriate subspaces of complex codimension one in $\MM_n
\times R^{9-2n,1}$\cite{GP}.
Since the D7-branes carry 1 unit of RR charge, the
monodromy around a D7-brane is $T$.
We must place these D-branes in such a way that for any $S^2$ embedded
in $\MM_n$, the total monodromy around all the points where this $S^2$
intersects the orientifold plane and the D-branes vanish (see Fig.1).
Also, the D-brane configuration must be such so as not to break any
further supersymmetry.
We shall see later how this can be achieved in practice.
\begin{figure}[!ht]
\begin{center}
\leavevmode
\epsfbox{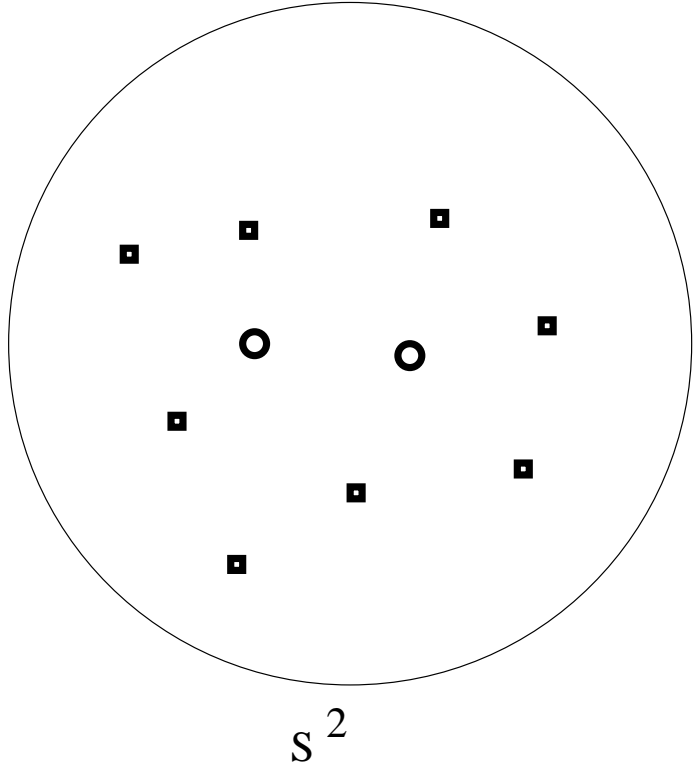}
\end{center}
\caption[]{\small
In this figure we have displayed a two dimensional
sphere $S^2$ embedded in $\MM_n$. The black squares represent
points where the D-branes intersect this sphere, and the black
circles represent points where the orientifold planes intersect
the sphere. The monodromy around a curve enclosing all the 
squares and the circles must be trivial since this curve can
be contracted to a point in $S^2$. This forces the number of
circles to be even and the number of squares to be four times the
number of circles.}
\end{figure}

Let us now turn to a
brief review of F-theory\cite{FTHEORY}. The
starting point of an F-theory compactification 
is an elliptically fibered (fibers are two dimensional tori)
Calabi-Yau $(n+1)$-fold 
$\MM_{n+1}$ on a base $\BB_n$.
Let $\vu$ be the complex coordinates on $\BB_n$, and
$\tau(\vu)$ denote the modular parameter of the fiber $T^2$ as a
function of $\vu$. 
By definition, F-theory compactified
on $\MM_{n+1}$ is type IIB on $\BB_n\times R^{9-2n,1}$ with
\be \label{e8}
\lambda(\vu) = \tau(\vu)\, .
\ee
It will be useful to consider the
Weierstrass form of elliptically fibered manifold:
\be \label{e9}
y^2 = x^3 + f(\vu) x + g(\vu)\, ,
\ee
where $x$, $y$ are complex variables, and 
$f(\vu)$, $g(\vu)$ are sections of appropriate line bundles on
$\BB_n$. In particular, we shall choose $f$ and $g$ to be
sections of $L^{\otimes 4}$ and $L^{\otimes 6}$ 
respectively, where
$L^{\otimes n}$ denote the $n$th power of some line bundle $L$.
We can then make sense of eq.\refb{e9} by regarding $x$ and $y$
as elements of $L^{\otimes 2}$ and $L^{\otimes 3}$
respectively.
For every $\vu$ we have a torus labelled by $(x,y)$ satisfying
\refb{e9},
with modular parameter $\tau$ given by:
\be \label{e10}
j(\tau) = 4 \cdot (24f)^3/(4 f^3 + 27 g^2)\, .
\ee
$j$ is the modular function with a single pole at $i\infty$.
F-theory on this elliptically fibered manifold is type IIB on $\BB_n$ with
\be \label{e11}
j(\lambda(\vu)) = 4 \cdot (24f)^3/(4 f^3 + 27 g^2)\, .
\ee  
$j(\lambda)\to \infty$ at zeroes of 
\be \label{e12}
\Delta \equiv (4f^3 + 27 g^2)\, .
\ee
At these points
$\lambda\to i\infty$ {\it up to SL(2,Z) transformation}.
These are surfaces of complex codimension one, and are known as
the locations of the seven branes, although, as we shall see soon, they
are not necessarily Dirichlet seven branes.
Monodromy around each of these seven branes is conjugate to
SL(2,Z) transformation $T$.

We shall now take an appropriate `weak coupling limit' such that the
monodromy around the zeroes of $\Delta$ look identical to that of
an orientifold.
For this, let us take: 
\ben \label{e13}
f &=& - 3 h^2 + C \eta \nonumber \\
g &=& - 2 h^3 + C h \eta + C^2 \chi\, .
\een
Here $C$ is a constant, and
$h$, $\eta$, $\chi$ are sections of line bundles 
$L^{\otimes 2}$, $L^{\otimes 4}$ and $L^{\otimes 6}$
respectively. 
There is no loss of generality in the choice of $f$ and $g$ given
in \refb{e13}, since
for fixed $h$ and $C$, we can vary
$\eta$ and $\chi$ to span the whole range of $f$ and $g$. 
On the other hand,
there is clearly a redundancy in this choice, since
$C$ and $h$ could be absorbed in $\eta$ and $\chi$.
Put another way, 
for a given $f$ and $g$, we can choose
$C$, $\eta$, $h$ and $\chi$ in many different ways.
Nevertheless
we shall keep this redundancy, as this will help us take the weak
coupling limit properly.

With the above representation of $f$ and $g$, we get
\begin{eqnarray} \label{e14}
\Delta &=& (4f^3+27 g^2) \nonumber \\
&=& C^2 \{ \eta^2 (4 C\eta - 9h^2) + 54 h (C\eta - 2h^2)\chi
+ 27 C^2 \chi^2\} \, ,
\end{eqnarray}
\be \label{e15}
j(\lambda) = 4\cdot (24)^3 \cdot (C\eta - 3h^2)^3/\Delta\, .
\ee
Now we take the `weak coupling limit' $C\to 0$.
In this limit
\be \label{e16}
\Delta\simeq C^2 (-9 h^2) (\eta^2 + 12 h\chi)\, .
\ee
Thus the zeroes of $\Delta$ are at
\be \label{e17}
h=0, \qquad \hbox{and} \qquad \eta^2 + 12 h\chi =0\, .
\ee
Also, $j(\lambda)$ is large everywhere on the base except in regions
\be \label{e18}
| h|\sim |C|^{1/2}\, .
\ee
\begin{figure}[!ht]
\begin{center}
\leavevmode
\epsfbox{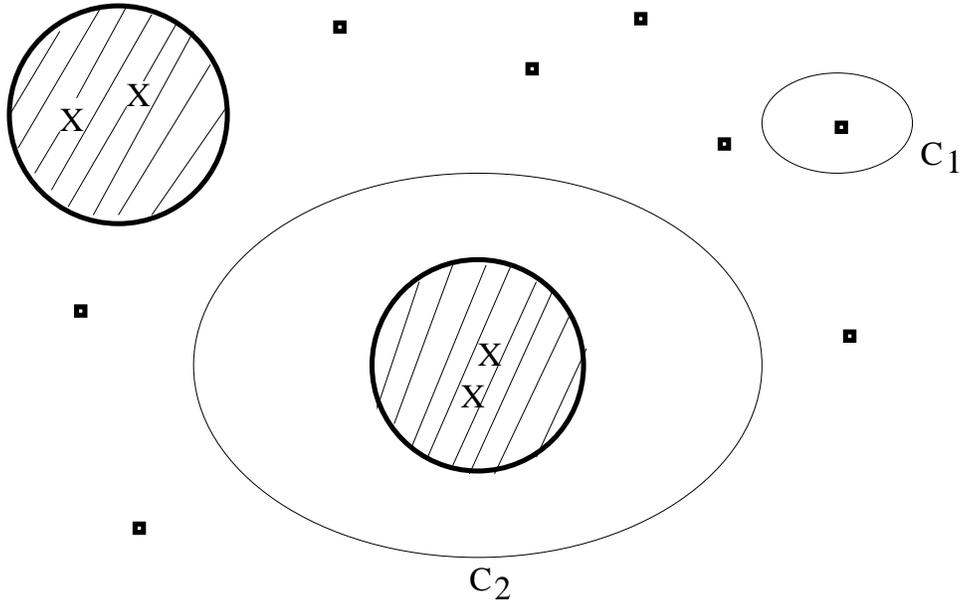}
\end{center}
\caption[]{\small
This figure displays a two dimensional section of
$\BB_n$. The black squares represent the zeroes of
$(\eta^2+12h\chi)$.
The shaded regions denote the region
$|h|\sim |C|^{1/2}$, and the two black crosses inside each of the shaded
region are the two zeroes of $\Delta$ near $h=0$. A contour
around only one of these black crosses must pass through the shaded
region, whereas a contour around both crosses can avoid the shaded
region. $C_2$ denotes such a contour. $C_1$ is a contour around a
zero of $(\eta^2+12h\chi)$. In the unshaded region of this diagram,
$Im(\lambda)$ is large. As shown in the text,
under F-theory $-$ orientifold correspondence,
each shaded region is mapped to a black circle in Fig.1.}
\end{figure}

Let us now
recall  that large $j(\lambda)$ corresponds to large $Im(\lambda)$ up
to an 
SL(2,Z) transformation. Thus, at every point in the region 
$|h|>>|C|^{1/2}$, either $Im(\lambda)$ is large, or
$\lambda$ approaches a rational point on the real axis. 
Since $h=0$ corresponds to a surface of
real codimension two, the region $h\ne 0$ is connected. Thus for
small $|C|$, the region $|h|>>|C|^{1/2}$, where $j(\lambda)$ is
large, is also connected (the unshaded region in Fig. 2). 
This shows that if we choose our
SL(2,Z) convention in such a way that $Im(\lambda)$ is large at
one point in the region $|h|>>|C|^{1/2}$ (which can always be done), 
then it must remain large in the whole of this region. This is the
way we shall choose our convention from now on.
Since large $Im(\lambda)$ corresponds to weak coupling, we see
that in this convention, the $C\to 0$ limit corresponds to
coupling constant being small in most part of the base. It is in this
sense that $C\to 0$ represents weak coupling limit.

We shall now compute the monodromy around the zeroes of $\Delta$
given in \refb{e16}.
In doing this computation, we should keep in mind that
as long as we choose a contour avoiding the $|h|\sim |C|^{1/2}$
regions ({\it e.g.} $C_1$ or $C_2$ in Fig.2),
$Im(\lambda)$ is large all along the contour, and hence the only
allowed SL(2,Z) monodromy along such a contour is $\pm T^n$.
First we shall focus on the monodromies around the 7-branes at
\be \label{e20}
\eta^2 + 12 h\chi =0\, .
\ee
Unless this expression has an accidental double zero,
this represents single zeroes of $\Delta$.  Thus
monodromy around such a singularity must be SL(2,Z) conjugate to $T$.
We can choose the contour around this hyperplane 
keeping it away from the $|h|\sim |C|^{1/2}$ region ({\it e.g.} $C_1$
in Fig.2). 
Hence, the monodromy must be of the form $\pm T^n$.
Combining these two requirements, we see that 
the monodromy must be  $T$.
Since this is the monodromy around a D7-brane, we see that for
small $|C|$,
\refb{e20} represent the locations of Dirichlet 7-branes.

Next we turn to analyze the 
monodromy around the hypersurface
\be \label{e21}
h=0\, .
\ee
Note that $\Delta$ has double zeroes at these locations.
However, this double zero appears only as $C\to 0$.
For small non-zero $C$, $\Delta$ will have a pair of zeroes around
$h=0$ as displayed in
Fig.2. The monodromies around these individual zeroes are
conjugate to $T$.
Let us take these monodromies  to be
\be \label{e22}
MTM^{-1} \qquad \hbox{and} \qquad NTN^{-1}\, ,
\ee
respectively.  Here $M$ and $N$ are two SL(2,Z) matrices.
Then the
total monodromy around the $h=0$ surface is 
\be \label{e23}
MTM^{-1}NTN^{-1}\, .
\ee
Since the contour around the individual zeroes of $\Delta$ around
$h=0$ must pass through the region $|h|\sim |C|^{1/2}$, we cannot
conclude that these monodromies must be of the form $\pm T^n$
(see Fig. 2).
However, a 
contour $C_2$ encircling both these zeroes can be taken to be
away 
from this region. Hence \refb{e23} must be of the form 
$\pm T^n$. We shall now try to determine the value of $n$, as
well as the overall sign.
For this, note that for large $Im(\lambda)$, 
\be \label{e24}
j(\lambda) \sim e^{-2\pi i\lambda}\, .
\ee
Thus
\be \label{e25}
n\equiv \ointop_{C_2} d\lambda = -{1\over 2\pi i} 
\ointop_{C_2} d \ln j(\lambda)\, .
\ee
Thus in order to calculate $n$, we
need to calculate the change in $\ln j(\lambda)$ as we go once
around the contour.
For small $|C|$ and $|h|>>|C|^{1/2}$, we have
\be \label{e26}
j(\lambda) \sim h^4/C^2(\eta^2 + 12 h \chi)\, .
\ee
Thus along a contour $C_2$ around $h=0$, $\ln j(\lambda)$ changes by
$4\cdot 2\pi i$. 
This gives, from \refb{e25}, $n=-4$. Hence the monodromy along the
contour $C_2$ is  $\pm T^{-4}$.

Next we turn to the
determination of the overall sign.
For this, 
recall that this monodromy must be expressible as $MTM^{-1}NTN^{-1}$
{\it i.e.} we need 
\be \label{e27}
MTM^{-1} NTN^{-1} = \pm T^{-4}\, .
\ee
It turns out that the most general solution of this equation is:
\be \label{e28}
MTM^{-1} = \pmatrix{1-p & p^2 \cr -1 & 1+p}\, , \qquad
NTN^{-1} = \pmatrix{-1-p & (p+2)^2 \cr -1 & 3+p}\, ,
\ee
giving
\be \label{e29}
MTM^{-1} NTN^{-1} = - T^{-4}\, .
\ee
Here $p$ is an arbitrary integer.
This shows that the monodromy around $h=0$ is  $-T^{-4}$. In
other words, for small $C$, the
$h=0$ plane behaves like an orientifold plane!

Thus we see that in the $C\to 0$ limit, the F-theory background can be
identified to that of an orientifold with,
\begin{enumerate}
\item
Orientifold 7-planes at
$h(\vu)=0$, and
\item Dirichlet 7-branes at
$\eta(\vu)^2 + 12 h(\vu) \chi(\vu) =0$.
\end{enumerate}
This analysis also shows that for
small but finite $C$,
the orientifold plane splits into two seven branes
lying close to the surface $h=0$.
This reflects a phenomenon already observed earlier in a much simpler
situation $-$ namely orientifold of type IIB compactified on a
two dimensional torus\cite{SEF}.

We would also like to find
the original manifold whose orientifold this theory is.
It is clear that this manifold
must be a double cover of the base $\BB_n$, branched along the 
orientifold plane $h(\vu)=0$.
Let us now consider the manifold:
\be \label{e30}
\MM_n: \quad\xi^2 = h(\vu)\, ,
\ee
where
$\xi$ is an element of the line bundle $L$.
This manifold has a $Z_2$ isometry 
\be \label{e31}
\sigma: \qquad \xi\to -\xi\, .
\ee
Fixed point set under this isometry corresponds to the
complex codimension one submanifold $\xi=0$. Using eq.\refb{e30},
this gives $h(\vu)=0$.
Since for every point $\vu$
in $\BB_n$, except those at $h(\vu)=0$,
we have two points in $\MM_n$ given by $(\vu,
\xi=\pm\sqrt{h(\vu)})$, $\MM_n$ is a double cover of $\BB_n$ branched
along $h(\vu)=0$.
Thus $\BB_n$ can be identified with $\MM_n/\sigma$. 
This, in turn, shows that the precise description of the
orientifold that we have found is
\be \label{e32}
\hbox{Type IIB on} \, \,  \MM_n
\times R^{9-2n,1}/(-1)^{F_L}\cdot\Omega\cdot\sigma\, .
\ee

The next question that arises naturally is:
does $\MM_n$ represent a Calabi-Yau $n$-fold?
To answer this question, let us consider the original
Calabi-Yau manifold $\MM_{n+1}$ described in eq.\refb{e9} with
$f$ and $g$ being  sections of $L^{\otimes 4}$ and 
$L^{\otimes 6}$ respectively, and $x$ and $y$ being 
coordinates on $L^{\otimes 2}$ and $L^{\otimes 3}$ respectively. 
In order that $\MM_{n+1}$ is Calabi-Yau, we need its first Chern class to
vanish. This imposes the following restriction on $L$:
\be \label{e33}
c_1(\BB_n) + c_1(L) (3 + 2 - 6) = 0\, .
\ee
In the coefficient of $c_1(L)$ the factors 3 and 2 represent the
fact that $y$ and $x$ are coordinates on $L^{\otimes 3}$ and
$L^{\otimes 2}$ respectively, whereas the factor 6 represents that
the constraint \refb{e9} belongs to $L^{\otimes 6}$.
On the other hand, since $h$ is a section of $L^{\otimes 2}$, and
$\xi$ is a coordinate on $L$, in order that the auxiliary
manifold $\MM_n$ described in \refb{e30} is a Calabi-Yau
manifold, we must have:
\be \label{e35}
c_1(\BB_n) + c_1(L)(1-2) = 0\, .
\ee
But this is identical to the condition \refb{e33} for $\MM_{n+1}$ to
be Calabi-Yau. Thus we see that $\MM_n$ also describes a
Calabi-Yau manifold, provided it is a non-singular
manifold.

This finishes the outline of the general procedure by which we
can map an F-theory compactification to an orientifold of type
IIB in appropriate weak coupling limit. We shall now illustrate
this by means of a few examples. 
The first example we shall consider will be 
F-theory on Calabi-Yau 3-fold on base $CP^1\times CP^1$.
Let $u,v$ be the affine coordinates on $CP^1\times CP^1$. An
elliptically fibered Calabi-Yau 3-fold corresponds to  choosing 
$f(u,v)$ to be a polynomial of degree (8,8) in $(u,v)$
and $g(u,v)$ to be a polynomial of degree (12,12) in $(u,v)$ in
eq.\refb{e9}.
Then $h(u,v)$, $\eta(u,v)$ and $\chi(u,v)$ are respectively
polynomials of degree (4,4), (8,8) and (12,12) in $(u,v)$.
According to our analysis, in the weak coupling limit this describes
an orientifold 

\centerline{Type IIB on $\MM_2\times R^{5,1}
/(-1)^{F_L}\cdot\Omega\cdot\sigma$}

\noindent where $\MM_2$ corresponds to the manifold
\be \label{ep1}
\MM_2: \qquad \xi^2=h(u,v)\, .
\ee
This corresponds to a K3 surface. It can be shown that
$\sigma$  ($\xi\to -\xi$) describes a Nikulin involution\cite{NIKULIN}
on this surface with 
\be \label{ep2}
(r,a,\delta)=(2,2,0)\, .
\ee
The D-branes are situated at
\be \label{ep3}
\eta(u,v)^2 + 12 h(u,v)\chi(u,v) = 0\, .
\ee
We can simplify this model further by going to the orbifold limit
of this K3.  For this, we choose:
\ben \label{ep4}
h(u,v) &=& \prod_{\alpha=1}^4 (u -\tu_\alpha) (v - \tv_\alpha)
\nonumber \\
\eta(u,v) &=& \prod_{i=1}^8 (u -u_i) (v - v_i), \qquad \chi=0\, .
\een
Here
$\tu_\alpha$, $\tv_\alpha$, $u_i$, $v_i$ are constants.
This gives the following defining equation for $\MM_2$:
\be \label{ep5}
\MM_2: \quad \xi^2 =
\prod_{\alpha=1}^4 (u -\tu_\alpha) (v - \tv_\alpha)\, .
\ee
This corresponds to the $T^4/Z_2$ orbifold limit of K3.
The $D7$-branes are now located in pairs at:
\be \label{ep6}
u=u_i \qquad \hbox{and at} \qquad v=v_i.
\ee
This model can be shown\cite{GIMON2} to be
T-dual to the Gimon-Polchinski model\cite{SAGN,GP}.

By using similar methods one can show that F-theory on Calabi-Yau
3-folds on base $F_1$ and $F_4$ can be mapped to 
type IIB on 
$K3\times R^{5,1}/(-1)^{F_L}\cdot\Omega\cdot\sigma$ where $\sigma$
corresponds to the Nikulin involution:
\ben \label{ep7}
(r,a,\delta)&=&(2,2,1) \qquad \hbox{for} \qquad \hbox{base} \quad F_1
\nonumber \\
(r,a,\delta)&=&(2,0,0) \qquad \hbox{for} \qquad \hbox{base} \quad
F_4\, .
\een

Using mirror symmetry, one can further relate type IIB on 
$K3/(-1)^{F_L}\cdot\Omega\cdot\sigma$ to type I on 
a mirror $K3$\cite{SETHI} which we shall denote by $K3'$.
The three different choices of $\sigma$ correspond to three different
choices of gauge bundles for type I on $K3'$.
In order to give a more specific description of this
correspondence, let us 
take a two sphere $C$ inside $K3'$, and let 
$C_N$ and $C_S$ denote the northern and southern hemispheres of
$C$. Also let
$g_N$  and $g_S$ denote the holonomies along the boundaries of $C_N$  and
$C_S$ respectively in the {\it vector representation} of SO(32).
Then $g_N g_S^{-1}$ is unity when regarded as an element of the
group $Spin(32)/Z_2$, but not necessarily as an element of the
group $SO(32)$.
We
define $\wt w_2$ to be an element of the second homology class of
$K3'$ such that:
\be \label{ep9}
g_N g_S^{-1} = \exp\big(i\pi (\wt w_2 \cap C)\big)\, ,
\ee
where $(\wt w_2\cap C)$ denotes the intersection number of $\wt
w_2$ and $C$.
Then the gauge bundle of type I on $K3'$ can belong to either of the
three classes characterized by the following properties on $\wt
w_2$\cite{WITSIX,ASPINEW}:
\begin{itemize}
\item
$\wt w_2=0$, 
\item 
$(\wt w_2 \cap \wt w_2)=0 \quad mod \quad 4$,
\item
$(\wt w_2 \cap \wt w_2)=2 \quad mod \quad 4$.
\end{itemize}
\begin{figure}[!ht]
\begin{center}
\leavevmode
\epsfbox{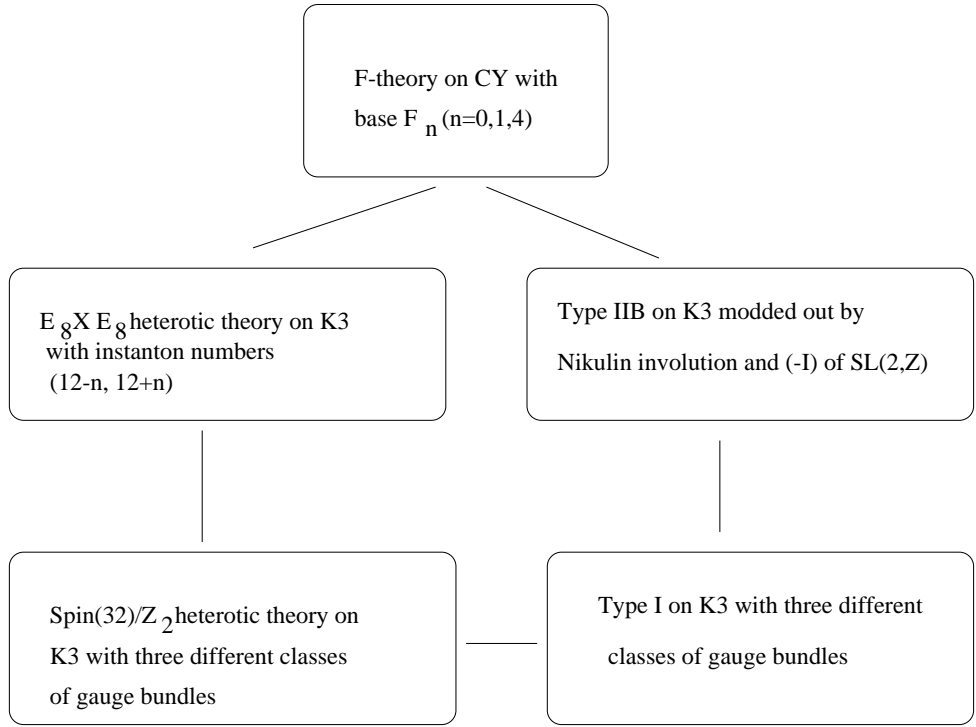}
\end{center}
\caption[]{\small The Duality Cycle}
\end{figure}

One can show that\cite{SETHI} these three 
different classes of type I
compactifications are related, by a 
mirror transformation, to
type IIB on $(K3\times R^{5,1}/(-1)^{F_L}\cdot\Omega\cdot \sigma)$
where
$\sigma$ is a Nikulin involution with 
\begin{itemize}
\item
$(r,a,\delta)=(2,0,0)$  for $\wt w_2=0$,
\item
$(r,a,\delta)=(2,2,1)$  for $(\wt w_2\cap \wt w_2)=2$ mod
4,
\item
$(r,a,\delta)=(2,2,0)$  for $(\wt w_2\cap\wt w_2)=0$ mod 4.
\end{itemize}
As we have already described in detail earlier, these three
theories are in turn related to F-theory on Calabi-Yau three
folds on base $F_4$, $F_1$ and $F_0(\equiv CP^1\times CP^1)$
respectively.

The results described in this talk, when combined with the other
known duality symmetries of string theory, leads us to the
duality cycle displayed in Fig.3 which was first guessed by Gimon and
Johnson\cite{GIMJOH}. There exists independent understanding of
the duality
involving each link in this cycle. In this talk we discussed two
of these links, the ones connecting the topmost box with the two
boxes on the right hand side of the diagram. Duality between
F-theory on Calabi-Yau with base $F_n$ and $E_8\times E_8$
heterotic string theory on K3 with instanton number $(12-n)$ in
the first $E_8$ and $(12+n)$ in the second $E_8$ was discussed in
ref.\cite{FTHEORY}. The equivalence between $E_8\times E_8$
theory on $K3$ and $Spin(32)/Z_2$ hetertotic theory on $K3$ was
established in refs.\cite{WITSIX,ASPINEW}. Finally the horizontal
link in the lower part of the diagram follows from the
conjectured duality between type I and SO(32) heterotic string
theory in ten dimensions\cite{WITOLD}.

\end{document}